\documentclass[pre,showpacs,superscriptaddress,eqsecnum,twocolumn]{revtex4}

\usepackage{amsmath}
\usepackage{amsfonts}
\usepackage{amssymb}
\usepackage{graphicx}
\usepackage{siunitx}

\begin{document}

\renewcommand{\thefootnote}{\roman{footnote}}

\title{Nonequilibrium fluctuation-induced Casimir pressures in liquid mixtures}

\author{T. R. Kirkpatrick}
\affiliation{Institute for Physical Science and Technology, University of Maryland, College Park, Maryland 20877, USA}
\affiliation{Department of Physics, University of Maryland, College Park, Maryland 20877, USA}

\author{J. M. Ortiz de Z\'arate}
\affiliation{Departamento de F\'{\i}sica Aplicada I, Facultad de F\'{\i}sica, Universidad Complutense, 28040 Madrid, Spain}

\author{J. V. Sengers}
\email{sengers@umd.edu}
\affiliation{Institute for Physical Science and Technology, University of Maryland, College Park, Maryland 20877, USA}

\date{\today}

\begin{abstract}
In this article we derive expressions for Casimir-like pressures induced by nonequilibrium concentration fluctuations in liquid mixtures. The results are then applied to liquid mixtures in which the concentration gradient results from a temperature gradient through the Soret effect. A comparison is made between the pressures induced by nonequilibrium concentration fluctuations in liquid mixtures and those induced by nonequilibrium temperature fluctuations in one-component fluids. Some suggestions for experimental verification procedures are also presented.
\end{abstract}


\maketitle

\section{INTRODUCTION\label{S1}}

Fluctuation-induced forces will appear in confined fluids when long-ranged fluctuations are present~\cite{KardarGolestanian}. These phenomena are also frequently referred to as Casimir effects. A well-known example is the Casimir effect in critical systems, where the forces are induced by long-range critical fluctuations~\cite{KrechBook,Krech1,GambassiEtAl1}. A more recent example is the nonequilibrium Casimir effect in fluids. Thermal fluctuations in fluids in nonequilibrium steady states are large and very long ranged~\cite{DorfmanKirkpatrickSengers,BelitzKirkpatrickVotja}. These nonequilibrium fluctuations are particularly spectacular in fluids in the presence of a temperature or concentration gradient. They arise from a coupling between the heat-diffusion or mass-diffusion mode and the viscous mode through the convective term in the fluctuating hydrodynamics equations~\cite{BOOK,miChaperRSC2}. As a consequence, they induce Casimir-like forces much larger than fluctuation-induced forces in fluids in thermodynamic equilibrium.

In some previous works, fluctuation-induced forces in one-component fluids in the presence of a temperature gradient have been considered~\cite{miPRL2,miPRE2014,AminovEtAl,miPRE2015}. The purpose of the present paper is to study fluctuation-induced forces in liquid mixtures in the presence of a concentration gradient. Such a concentration gradient may either be isothermal~\cite{GiglioNature,VailatiGiglio3}, or it may be the result of a temperature gradient through the Soret effect~\cite{SegreEtAlPRE,Mixtures3,VailatiGiglio1,miJCP,miDynamics14}. A brief account of our principal results has been presented in a recent letter~\cite{miCasimirBin}. In the present paper we study this nonequilibrium (NE) Casimir effect in liquid mixtures in more detail.

We shall proceed as follows. In Section~\ref{S2} we derive the relationship between the NE fluctuation-induced pressure and the NE concentration fluctuations. In dealing with the NE fluctuations in liquid mixtures we shall use a frequently adopted large-Lewis-number approximation, in which concentration fluctuations and temperature fluctuations decouple. In Section~\ref{S3} we consider the intensity of the NE concentration fluctuations deduced from NE fluctuating hydrodynamics. The resulting expressions for the NE fluctuation-induced pressure will be investigated in Section~\ref{S4}. We conclude with a discussion of the results in Section~\ref{S5}. In the Appendix we relate the pressure expansion equation for the fluctuation-induced pressure derived in Section~\ref{S2} to the usual statistical mechanical definition of the pressure.

\section{RELATION BETWEEN NE PRESSURE AND NE CONCENTRATION FLUCTUATIONS\label{S2}}

To derive the expression for a nonequilibrium fluctuation-induced pressure, we consider the pressure as a function of the fluctuating conserved densities, which for a liquid mixture are the fluctuating energy density $e+\delta e$, the fluctuating mass density $\rho _{1} +\delta \rho _{1}$ of component 1 (solute) and the fluctuating mass density $\rho _{2} +\delta \rho _{2}$ of component 2 (solvent):
\begin{equation} \label{GrindEQ__2_1_}
p\left(e+\delta e,\rho _{1} +\delta \rho _{1} ,\rho _{2} +\delta \rho _{2} \right)=p\left(e,\rho _{1} ,\rho _{2} \right)+\delta p.
\end{equation}
We expand $p\left(e+\delta e,\rho _{1} +\delta \rho _{1} ,\rho _{2} +\delta \rho _{2} \right)$ in a Taylor series in terms of $\delta e$, $\delta \rho _{1}$, and $\delta\rho _{2}$. Dealing with the slow mass-diffusion mode, we can neglect the fast propagating sound modes, and, hence the linear fluctuation contribution to the pressure:
\begin{equation} \label{GrindEQ__2_2_}
\left(\frac{\partial p}{\partial e} \right)_{\rho _{1} ,\rho _{2} }{\mkern-7mu}\delta e+\left(\frac{\partial p}{\partial \rho _{1} } \right)_{e,\rho _{2} }{\mkern-7mu} \delta \rho _{1} +\left(\frac{\partial p}{\partial \rho _{2} } \right)_{e,\rho _{} }{\mkern-7mu} \delta \rho _{2} =0.
\end{equation}
Retaining only terms quadratic in the fluctuations, we thus obtain for the Taylor expansion of $\delta p$:
\begin{equation} \label{GrindEQ__2_3_}
\delta p=\frac{1}{2} \left[\begin{array}{l} \left(\dfrac{\partial ^{2} p}{\partial e^{2} } \right)_{\rho _{1} ,\rho _{2} }{\mkern-11mu}\left(\delta e\right)^{2} +\left(\dfrac{\partial ^{2} p}{\partial \rho _{1}^{2} } \right)_{e,\rho _{2} }{\mkern-11mu}\left(\delta \rho _{1} \right)^{2} \\[14pt]+ \left(\dfrac{\partial ^{2} p}{\partial \rho _{2}^{2} } \right)_{e,\rho_1}{\mkern-11mu} \left(\delta \rho _{2} \right)^{2} +  2\left(\dfrac{\partial ^{2} p}{\partial e\partial \rho _{1} } \right)\delta e\delta \rho _{1}\\[14pt] +2\left(\dfrac{\partial ^{2} p}{\partial e\partial \rho _{2} } \right)\delta e\delta \rho _{2} +2\left(\dfrac{\partial ^{2} p}{\partial \rho _{1} \partial \rho _{2} } \right)\delta \rho _{1} \delta \rho _{2} \end{array}\right].
\end{equation}
In liquid mixtures there are two diffusion modes that are linear combination of heat diffusion and mass diffusion~\cite{Mixtures3,Wood}. An important parameter for mixtures is the Lewis number $Le=D_{T} /D$, which is the ratio of the thermal diffusivity $D_{T}$ and the mass-diffusion coefficient $D$. For liquid mixtures the Lewis number is substantially larger than unity. Hence, in dealing with fluctuations in liquid mixtures one often adopts a large-Lewis-number approximation~\cite{BOOK}. When $Le\gg 1$, the two diffusion modes decouple into a pure temperature fluctuation mode with a decay time proportional to $D_{T}$ and a concentration fluctuation mode with a decay time proportional to $D$~\cite{VelardeSchechter}. Hence, to get the slowest-mode contribution when $Le\gg 1$, we not only may neglect linear pressure fluctuations in accordance with Eq. \eqref{GrindEQ__2_2_}, but also temperature fluctuations:
\begin{equation} \label{GrindEQ__2_4_}
\delta T=0.
\end{equation}
For the concentration variable we adopt the mass fraction $w=\rho _{1} /\rho$ of the solute, where $\rho =\rho _{1} +\rho_{2}$ is the mass density of the mixture. From Eqs.~\eqref{GrindEQ__2_2_} and~\eqref{GrindEQ__2_4_} it follows that the fluctuations $\delta e$, $\delta \rho_{1}$, and $\delta \rho _{2}$ can be related to the concentration fluctuation $\delta w$ at constant pressure $p$ and constant temperature $T$:
\begin{align} \label{GrindEQ__2_5_}
\delta e&=\left(\frac{\partial e}{\partial w} \right)_{\mkern-7mu{p,T}} \delta w,& \delta \rho _{1}&=\left(\frac{\partial \rho _{1} }{\partial w} \right)_{\mkern-7mu{p,T}} \delta w,\\
\delta \rho _{2}&=\left(\frac{\partial \rho _{2} }{\partial w} \right)_{\mkern-7mu{p,T}} \delta w.\notag
\end{align}
Substituting Eq.~\eqref{GrindEQ__2_5_} into Eq.~\eqref{GrindEQ__2_3_} and taking an average, we obtain for the fluctuation-induced pressure $p_{{\rm NE}}^{w} =\left\langle \delta p\right\rangle $:
\begin{equation} \label{GrindEQ__2_6_}
p_{{\rm NE}}^{w} \left({\bf r}\right)=\frac{1}{2} A_{w} \left\langle \left[\delta w\left({\bf r}\right)\right]^{2} \right\rangle _{{\rm NE}}
\end{equation}
with
\begin{equation} \label{GrindEQ__2_7_}
\begin{split} A_{w} &=\left(\dfrac{\partial ^{2} p}{\partial e^{2} } \right)_{\mkern-7mu{\rho _{1},\rho _{2}}} \left(\dfrac{\partial e}{\partial w} \right)_{\mkern-7mu{p,T}}^{2} +\left(\dfrac{\partial ^{2} p}{\partial \rho _{1}^{2} } \right)_{e,\rho _{2} } \left(\dfrac{\partial \rho _{1} }{\partial w} \right)_{p,T}^{2}\\& +\left(\dfrac{\partial ^{2} p}{\partial \rho _{2}^{2} } \right)_{e,\rho_1} \left(\dfrac{\partial \rho _{2} }{\partial w} \right)_{\mkern-7mu{p,T}}^{2}\\ & +2\left(\dfrac{\partial ^{2} p}{\partial e\partial \rho _{1} } \right)\left(\dfrac{\partial e}{\partial w} \right)_{\mkern-7mu{p,T}} \left(\dfrac{\partial \rho _{1} }{\partial w} \right)_{\mkern-7mu{p,T}} \\ &+2\left(\dfrac{\partial ^{2} p}{\partial e\partial \rho _{2} } \right)\left(\dfrac{\partial e}{\partial w} \right)_{p,T} \left(\dfrac{\partial \rho _{2} }{\partial w} \right)_{p,T} \\ &+ {2\left(\dfrac{\partial ^{2} p}{\partial \rho _{1} \partial \rho _{2} } \right)\left(\dfrac{\partial \rho _{1} }{\partial w} \right)_{p,T} \left(\dfrac{\partial \rho _{2} }{\partial w} \right)_{p,T} }. \end{split}\raisetag{40pt}
\end{equation}
The superscript $w$ indicates that $p_{{\rm NE}}^{w}$ is a pressure induced by concentration fluctuations. Only the NE concentration fluctuations $\left\langle \left[\delta w\left({\bf r}\right)\right]^{2} \right\rangle _{{\rm NE}}$ cause a renormalized pressure, since the equilibrium concentration fluctuations are already incorporated in the unrenormalized pressure. Just as for the case of a one-component fluid~\cite{miPRE2014}, our approach for deriving Eq.~\eqref{GrindEQ__2_6_} from an expansion of the pressure fluctuations in terms of the conserved quantities can be justified from an explicit mode-coupling theory generalized to nonequilibrium steady states, as shown in the Appendix. We comment that at this stage the fluctuation-induced pressure at any location ${\bf r}$ is related to the intensity of the NE concentration fluctuations at the same location. We shall see later that mechanical equilibrium requires that the actual fluctuation-induced pressure in a nonequilibrium steady state will be spatially uniform.

By using Eq.~\eqref{GrindEQ__2_2_} and noting that $d\rho _{1} =wd\rho +\rho~dw$ and $d\rho _{2} =\left(1-w\right)d\rho -\rho~dw$, Eq.~\eqref{GrindEQ__2_7_} can be simplified to
\begin{equation} \label{GrindEQ__2_8_}
\begin{split}
A_{w}=-\left(\frac{\partial p}{\partial e} \right)_{\rho ,w} & \left[\left(\frac{\partial ^{2} e}{\partial w^{2} } \right)_{p,T} -\left(\frac{\partial e}{\partial \rho } \right)_{p,w} \left(\frac{\partial ^{2} \rho }{\partial w^{2} } \right)_{p,T}\right. \\ &- \left. \frac{2}{\rho } \left(\frac{\partial e}{\partial w} \right)_{p,\rho } \left(\frac{\partial \rho }{\partial w} \right)_{p,T} \right].
\end{split}\raisetag{24pt}
\end{equation}

To evaluate Eq.~\eqref{GrindEQ__2_8_} we first note that $\left(\partial p/\partial e\right)_{\rho ,w} =\left(\gamma -1\right)/\alpha T$, where $\gamma =c_{p,w} /c_{V,w}$ is the ratio of the specific isobaric and isochoric heat capacities, and where $\alpha =-\rho ^{-1} \left(\partial \rho /\partial T\right)_{p,w}$ is the thermal expansion coefficient of the mixture at constant composition~\cite{ErnstHaugeVanLeeuwen}. The remainder of Eq.~\eqref{GrindEQ__2_8_} can be expressed in terms of the thermodynamic variables $w,p,T$ by using thermodynamic relations collected for mixtures by Wood~\cite{Wood}. Realizing that the thermodynamic field conjugate to the mass fraction $w$ is the difference $\mu$ between the chemical potentials of the solute and the solvent, we then obtain
\begin{equation} \label{GrindEQ__2_9_}
\begin{split}
A_{w} =-\frac{\rho \left(\gamma -1\right)}{\alpha T} &\left[\chi ^{-1} -T\left(\frac{\partial \chi ^{-1} }{\partial T} \right)_{p,w} \right. \\
&- \left.\frac{\rho c_{p,w} }{\alpha } \left(\frac{\partial \chi ^{-1} }{\partial p} \right)_{T,w} \right],
\end{split}
\end{equation}
where $\chi =\left(\partial w/\partial \mu \right)_{p,T}$ is an osmotic susceptibility. We note that in thermodynamic equilibrium $\left\langle \left(\delta w\right)^{2} \right\rangle _{{\rm E}} =k_{{\rm B}} T(\rho{V})^{-1} \chi $, where $k_{{\rm B}} $ is Boltzmann's constant and $V$ the volume of the system~\cite{BernePecora,MountainDeutch69,CohenEtAl}. The inverse osmotic susceptibility $\chi^{-1}$ can be related to the excess molar Gibbs energy $G^{\text{E}}$~\cite{LiThesis}:
\begin{align} \label{GrindEQ__2_10_}
\chi ^{-1} &=\left(\frac{\partial \mu }{\partial w} \right)_{p,T}\\
&=\frac{RT\, M}{w\left(1-w\right)M_{1} M_{2} } \left\{1-x_{1} x_{2} \left(\frac{\partial ^{2} G^{{\rm E}} /RT}{\partial x_{1} \partial x_{2} } \right)_{p,T} \right\},\notag
\end{align}
where $R$ is the molar gas constant, $x_{1}$ and $x_{2}$ are the mole fractions, and $M_{1}$ and $M_{2}$ the molar weights of solute and solvent, respectively, while $M=M_{1} x_{1} +M_{2} x_{2}$ is the molar weight of the mixture. Substitution of Eq.~\eqref{GrindEQ__2_10_} into Eq.~\eqref{GrindEQ__2_9_} yields finally
\begin{align} \label{GrindEQ__2_11_}
A_{w} =-\frac{\rho \left(\gamma -1\right)}{\alpha T} \frac{M^{3} }{M_{1}^{2} M_{2}^{2} } &\left[\left(\frac{\partial ^{2} H^{{\rm E}} }{\partial x_{1}^{2} } \right)_{p,T}
\right.\\&-\left.\frac{\rho c_{p,w} }{\alpha } \left(\frac{\partial ^{2} V^{{\rm E}} }{\partial x_{1}^{2} } \right)_{p,T} \right],\notag
\end{align}
to be substituted into Eq. \eqref{GrindEQ__2_6_}, where $H^{{\rm E}} $ is the excess molar enthalpy and $V^{{\rm E}}$ the excess molar volume. It is interesting to compare this result with the expression for the NE pressure $p_{{\rm NE}}^{T} $ induced by temperature fluctuations in a one-component fluid~\cite{miPRE2015}
\begin{equation} \label{GrindEQ__2_12_}
p_{{\rm NE}}^{T} \left({\bf r}\right)=\frac{1}{2} A_{T} \left\langle \left[\delta T\left({\bf r}\right)\right]^{2} \right\rangle _{{\rm NE}}
\end{equation}
with
\begin{equation} \label{GrindEQ__2_13_}
A_{T} =-\frac{\rho \left(\gamma -1\right)}{\alpha T} \left[\left(\frac{\partial ^{2} h}{\partial T^{2} } \right)_{p} -\frac{\rho c_{p} }{\alpha } \left(\frac{\partial ^{2} v}{\partial T^{2} } \right)_{p} \right],
\end{equation}
where $h$ is the specific enthalpy and $v$ the specific volume.

\section{NE CONCENTRATION FLUCTUATIONS\label{S3}}

We consider a liquid mixture between two horizontal plates separated by a distance $L$. We take a coordinate system with the $z$ axis in the vertical direction. The plates are located at $z=-L/2$ and at $z=+L/2$. The liquid mixture is subjected to a stationary concentration gradient ${\bf \nabla }w_{0}$, where $w_{0} \left(z\right)$ is the local average concentration which is assumed to be a linear function of $z$. We also assume that the liquid mixture is in a quiet mechanically stable state far away from any convective instability~\cite{SchechterEtAl,PlattenLegros}. Under these conditions, the NE concentration fluctuation $\delta w=\delta w\left({\bf r},t\right)$, which depends on the location ${\bf r}$ and the time $t$, satisfies a simple linearized fluctuating mass-diffusion equation:
\begin{equation} \label{GrindEQ__3_1_}
\rho \left[\frac{\partial \delta w}{\partial t} +\delta {\bf v}\cdot \nabla w_{0} \right]=\rho D\nabla ^{2} \delta w-\nabla \cdot \delta {\bf J},
\end{equation}
where $\delta {\bf J}$ is a fluctuating mass-diffusion flux~\cite{miChaperRSC2,SegreEtAlPRE,LawNieuwoudt}. This fluctuating mass-diffusion equation differs from the one in equilibrium by the presence of the term $\delta {\bf v}\cdot \nabla w_{0} $, which causes a coupling of the concentration fluctuations $\delta{w}$ with the velocity fluctuations $\delta\mathbf{v}$. The velocity fluctuations are to be determined from a fluctuating Stokes equation:
\begin{equation} \label{GrindEQ__3_2_}
\rho \frac{\partial \delta {\bf v\; }}{\partial t} =\eta \nabla ^{2} \delta {\bf v}-{\bf \nabla }\cdot \delta {\bf \Pi },
\end{equation}
where $\eta $ is the shear viscosity and $\delta {\bf \Pi }$ a fluctuating stress tensor. In fluctuating hydrodynamics  $\delta {\bf J}$ and $\delta {\bf \Pi }$ are assumed to satisfy a local fluctuation-dissipation theorem such that~\cite{BOOK,CohenEtAl,LawNieuwoudt,Foch1,miChaperRSC1}
\begin{multline} \label{GrindEQ__3_3_}
\left\langle \delta J_{i} \left({\bf r},t\right)\; \delta J_{j} \left({\bf r}',t'\right)\right\rangle =2k_{{\rm B}} T\rho\chi D~\delta_{ij} \\ \times \delta \left({\bf r}-{\it \; }{\it r}'\right)\delta \left(t-t'\right)
\end{multline}
and
\begin{multline} \label{GrindEQ__3_4_}
\left\langle \delta {\it \Pi }_{ij} \left({\bf r},t\right)\; \delta {\it \Pi }_{kl} \left({\bf r}',t'\right)\right\rangle =2k_{{\rm B}} T\, \eta \left(\delta _{ik} \delta _{jl} +\delta _{il} \delta _{jk} \right)\\ \times \delta \left({\bf r}-{\bf r}'\right)\delta \left(t-t'\right).
\end{multline}

Not only the Lewis number, but also the Schmidt number $Sc=\nu /D$, where $\nu =\eta /\rho $ is the kinematic viscosity, is commonly much larger than unity. It means that the viscous fluctuations decay much faster than the concentration fluctuations. Hence, for $Sc\gg 1$, we may neglect the time derivative in Eq. \eqref{GrindEQ__3_2_}~\cite{miEPJCasimir}. In principle, the thermophysical properties in Eqs.~\eqref{GrindEQ__3_1_}-\eqref{GrindEQ__3_4_} may depend on the concentration (and on the temperature, if a temperature gradient is present). In practice we identify them with their values at the center of the fluid layer; this has been shown to be a very good approximation~\cite{LiThesis}.

\begin{figure}[t]
\begin{center}
\includegraphics[width=0.9\columnwidth]{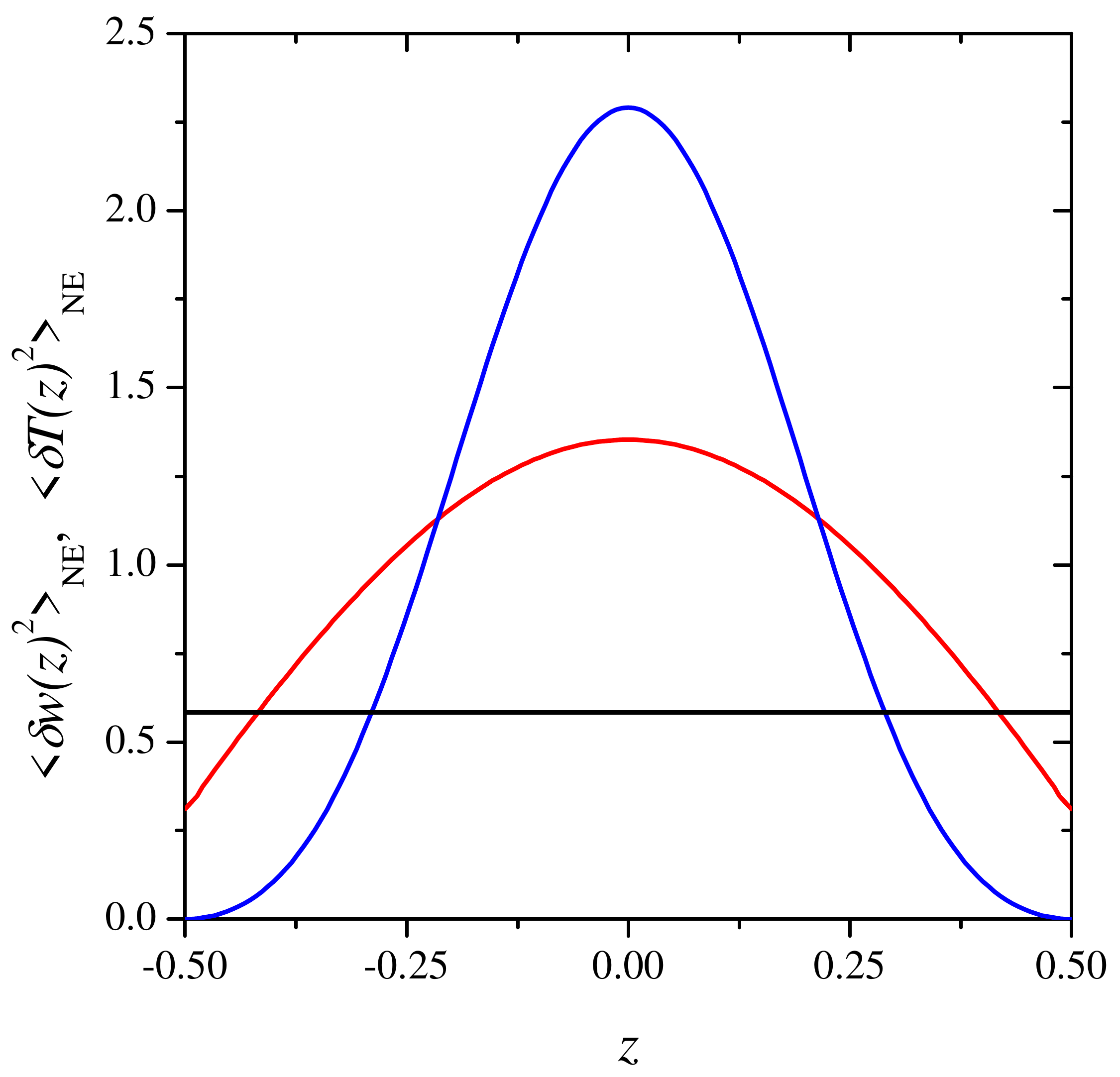}
\end{center}
\caption{(color on line). $\left\langle \left[\delta w\left(z\right)\right]^{2} \right\rangle _{{\rm NE}}$ as a function of $z$ obtained from Eq.~\eqref{GrindEQ__3_6_} with Eq.~\eqref{GrindEQ__3_13_} (red curve). For comparison we also show $\left\langle\left[\delta T\left(z\right)\right]^{2}\right\rangle_{{\rm NE}}$ for a one-component fluid in the presence of a temperature gradient obtained for rigid boundary conditions~\cite{miPRE2014} and infinite Prandtl number (blue curve). Both curves are normalized independently, so that the average value in the fluid layer for both types of fluctuations is unity. The horizontal line indicates the value of the $z$-independent approximation obtained by retaining only the constant $F_{0}$ in Eq.~\eqref{GrindEQ__3_13_} for $F(z)$, with the same normalization as the concentration fluctuations (red) curve~\cite{miCasimirBin}.}
\label{F1}
\end{figure}

These fluctuating hydrodynamics equations need to be solved subject to appropriate boundary conditions for the concentration and for the velocity fluctuations at the surfaces of the plates. Solutions have been presented in some previous publications, originally for artificial but mathematically convenient boundary conditions~\cite{Mexico} and, subsequently, for more realistic boundary conditions~\cite{miEPJCasimir,miIMT6}. Realistic boundary conditions are no-slip for the velocity fluctuations and impervious walls for the mass flow. If we neglect the effect of sound modes (divergence-free $\delta\mathbf{v}$), they are~\cite{miEPJCasimir,Chandra}:
\begin{align} \label{GrindEQ__3_5_}
\delta v_{z}& =\left(\frac{\partial \delta v_{z} }{\partial z} \right)=0,& \left(\frac{\partial \delta w}{\partial z} \right)&=0,&{\rm at}\quad z&=\pm \frac{L}{2} ,
\end{align}
where $\delta v_{z}$ is the fluctuation of the ${z}$ component of the velocity ${\bf v}$. For $Le\gg 1$ we have been able to obtain an explicit solution~\cite{miEPJCasimir} without a need for any Galerkin approximation that is usually considered for velocity fluctuations with rigid boundary conditions~\cite{BOOK}. From Ref.~\cite{miEPJCasimir} we find
\begin{equation} \label{GrindEQ__3_6_}
\left\langle \left[\delta w\left(z\right)\right]^{2} \right\rangle _{{\rm NE}} =\frac{k_{{\rm B}} T}{\rho \nu D} F(z)~L\left(\nabla w_{0} \right)^{2}
\end{equation}
with
\begin{equation} \label{GrindEQ__3_7_}
F(z)=\frac{1}{2\pi } \int _{0}^{\infty }q ~\widetilde{S} \left(q,z\right)dq,
\end{equation}
where $q$ is the magnitude of a dimensionless wave vector ${\bf q}_{\parallel } L$  associated with the concentration fluctuations in the x-y plane parallel to the plates. In Eq. \eqref{GrindEQ__3_7_} $\widetilde{S}\left(q,z\right)$ is a dimensionless generalized structure factor that contains two contributions:
\begin{equation} \label{GrindEQ__3_8_}
\widetilde{S}\left(q,z\right)=\widetilde{S}_{0} \left(q\right)+\widetilde{S}_{1} \left(q,z\right).
\end{equation}
The first term in Eq. \eqref{GrindEQ__3_8_} is a contribution independent of $z$ and represents the structure factor that is experimentally accessible in NE light-scattering or shadowgraph experiments:
\begin{equation} \label{GrindEQ__3_9_}
\widetilde{S}_{0} \left(q\right)=\frac{1}{q^{4} } +\frac{4\left(1-\cosh q\right)}{q^{5} \left(q+\sinh q\right)} .
\end{equation}
The second term is an additional $z$-dependent contribution:
\begin{align} \label{GrindEQ__3_10_}
\hspace*{-2pt}\widetilde{S}_{1} \left(q,z\right)&=2\sum _{\substack{N=0\\M=1}}^{\infty }\frac{\widetilde{A}_{NM} \cos \left(2N\pi\tilde{z}\right)\cos \left(2M\pi\tilde{z}\right)}{q^{2} +2N^{2} \pi ^{2} +2M^{2} \pi ^{2} }\\  & + {2\sum _{\substack{N=0\\M=0}}^{\infty }\frac{\widetilde{B}_{NM} \sin \left[\left(2N+1\right)\pi\tilde{z}\right]\sin \left[\left(2M+1\right)\pi\tilde{z}\right]}{q^{2} +\left[\left(2N+1\right)^{2} +\left(2M+1\right)^{2} \right]\dfrac{\pi^{2}}{2}}   } .\notag
 \end{align}
in terms of a dimensionless variable, $\tilde{z}=z/L$. The matrix elements in this double trigonometric series are given by
\begin{align} \label{GrindEQ__3_11_}
\widetilde{A}_{NM} \left(q\right)=&\frac{q^{2} \delta _{NM} }{\left(q^{2} +4N^{2} \pi ^{2} \right)^{2} }\\& + \frac{8q^{5} \left(1-\cosh q\right)\cos \left(N\pi \right)\cos \left(M\pi \right)}{\left(q+\sinh q\right)\left(q^{2} +4N^{2} \pi ^{2} \right)^{2} \left(q^{2} +4M^{2} \pi ^{2} \right)^{2} },\notag\\[24pt]
\widetilde{B}_{NM} \left(q\right)=&\frac{\pi^4 q^{2}\delta _{NM} }{\left[\dfrac{q^{2}}{\pi^2} +\left(2N+1\right)^{2}\right]^{2} }+\frac{8\pi^8 q^{5} }{q-\sinh q}\label{GrindEQ__3_12_}\\&{\mkern25mu}\times \frac{\left(1+\cosh q\right) \cos \left(N\pi \right)\cos \left(M\pi \right)}{\left[\dfrac{q^{2}}{\pi^2} +\left(2N+1\right)^{2}\right]^{2} \left[\dfrac{q^{2}}{\pi^2} +\left(2M+1\right)^{2}\right]^{2} }\notag
\end{align}
where $\delta _{NM}$ is a Kronecker delta. The integral obtained upon substituting Eq.~\eqref{GrindEQ__3_8_} into Eq.~\eqref{GrindEQ__3_7_} can be readily evaluated and we obtain
\begin{equation} \label{GrindEQ__3_13_}
F\left(z\right)=F_{0} +\frac{1}{2\pi } \int _{0}^{\infty }q~\widetilde{S}_{1}  \left(q,z\right)dq
\end{equation}
with
\begin{equation} \label{GrindEQ__3_14_}
F_{0} =\frac{1}{2\pi } \int _{0}^{\infty }q~\widetilde{S}_{0} \left(q\right) \simeq 3.11\times 10^{-3} .
\end{equation}
The contribution from the second term in Eq.~\eqref{GrindEQ__3_13_} is illustrated graphically in Fig.~\ref{F1}, where we show $\left\langle \left[\delta w\left(z\right)\right]^{2} \right\rangle _{{\rm NE}} $ as a function of $z/L$ (red curve) calculated numerically from Eq.~\eqref{GrindEQ__3_6_} and relative to the height-averaged value
\begin{equation} \label{GrindEQ__3_15_}
\begin{split}
\overline{\left\langle \left[\delta w\left(z\right)\right]^{2} \right\rangle }_{{\rm NE}}& =\frac{1}{L} \int _{-L/2}^{+L/2}\left\langle \left[\delta w\left(z\right)\right]^{2} \right\rangle _{{\rm NE}} dz\\& =\frac{k_{{\rm B}} T}{\rho \nu D} \overline{F}\; L\left(\nabla w_{0} \right)^{2} ,
\end{split}
\end{equation}
with
\begin{equation} \label{GrindEQ__3_16_}
\overline{F}= \int _{-1/2}^{+1/2}F(\tilde{z}) \; d\tilde{z}\simeq 5.724\times 10^{-3}.
\end{equation}

There are interesting differences between the profile for the intensity of the NE concentration fluctuations in a mixture and the profile for the intensity of the NE temperature fluctuations in a one-component fluid. For rigid boundaries the NE temperature fluctuations depend on the Prandtl number \textit{Pr}~\cite{EPJ}. The blue curve in Fig.~\ref{F1} shows the intensity of the NE temperature fluctuations in a one-component fluid, relative to the corresponding height-average~\cite{miPRE2014}, in the limit of infinite Prandtl number, consistent with the infinitely large Lewis and Schmidt numbers approximations adopted in the present paper. An important difference observed in Fig.~\ref{F1} is that in a fluid between two impervious thermally conducting walls the temperature fluctuations vanish at the walls, but the concentration fluctuations do not. Another difference is that the intensity of the NE temperature fluctuations approaches the walls with a vanishing slope, while the intensity of the NE concentration approaches the walls with a finite slope.

\section{NE PRESSURES INDUCED BY THE CONCENTRATION FLUCTUATIONS\label{S4}}

In principle the NE pressure induced by the concentration fluctuations is obtained by substituting Eq.~\eqref{GrindEQ__3_6_} into Eq.~\eqref{GrindEQ__2_6_}, so that
\begin{equation} \label{GrindEQ__4_1_}
p_{{\rm NE}}^{w} \left(z\right)=\frac{k_{{\rm B}} T}{2\rho \nu D} A_{w} F\left(z\right)L\left(\nabla w_{0} \right)^{2} .
\end{equation}
Equation \eqref{GrindEQ__4_1_} represents a fluctuation-induced pressure profile depending on the location $z$ in the liquid layer. Since the intensity of the NE concentration fluctuations does not vanish at the walls located at $z=\pm L/2$, we have emphasized in a previous publication that Eq.~\eqref{GrindEQ__4_1_} implies a direct fluctuation-induced pressure at the walls, in contrast to the fluctuation-induced pressure in a one-component fluid, where $p_{{\rm NE}}^{T} \left(z\right)$ vanishes at the walls~\cite{miCasimirBin}. We have realized subsequently that mechanical equilibrium requires that a pressure profile must cause a NE equilibrium contribution to the density profile $\rho _{{\rm NE}} \left(z\right)$, so as to establish a uniform NE pressure enhancement $\overline{p}_{{\rm NE}}^{w}$~\cite{miPRE2015}. The total pressure is then
\begin{equation} \label{GrindEQ__4_2_}
p=p_{{\rm eq}} +\overline{p}_{{\rm NE}}^{w} ,
\end{equation}
where $p_{{\rm eq}} $ is the equilibrium pressure. The fluctuation-induced NE density profile caused by $\overline{p}_{{\rm NE}}^{w} \left(z\right)$ is
\begin{equation} \label{GrindEQ__4_3_}
\rho _{{\rm NE}} \left(z\right)=-\left(\frac{\partial \rho }{\partial p} \right)_{T,w} \left[p_{{\rm NE}}^{w} \left(z\right)-\overline{p}_{{\rm NE}}^{w} \right].
\end{equation}
Conservation of mass requires that
\begin{equation} \label{GrindEQ__4_4_}
\overline{p}_{{\rm NE}}^{w} =\frac{1}{L} \int _{-L/2}^{+L/2}p_{{\rm NE}}^{w} \left(z\right)\; dz .
\end{equation}
Thus the uniform NE fluctuation-induced pressure is obtained by replacing $F\left(z\right)$ in Eq.~\eqref{GrindEQ__4_1_} with $\overline{F}$ from Eq.~\eqref{GrindEQ__3_16_}:
\begin{equation} \label{GrindEQ__4_5_}
\overline{p}_{{\rm NE}}^{w} =\frac{k_{{\rm B}} T}{2\rho \nu D} A_{w} \overline{F}\; L\left(\nabla w_{0} \right)^{2} .
\end{equation}
We note that for a fixed value of the concentration gradient the NE fluctuation-induced  pressure increases linearly with the distance $L$. This large NE Casimir effect is a direct consequence of the fact that in the absence of boundary conditions the intensity of the NE fluctuations varies with the wave number as $q^{-4} $, as can be seen from the leading term in Eq.~\eqref{GrindEQ__3_9_}.

A convenient experimental procedure for establishing a concentration gradient is by subjecting the liquid mixture to a stationary temperature gradient $\nabla T_{0} $. Then a concentration gradient will be established in the mixture through the Soret effect:
\begin{equation} \label{GrindEQ__4_6_}
\nabla w_{0} =-S_{T} w_{0} \left(1-w_{0} \right)\nabla T_{0} ,
\end{equation}
where $S_{T}$ is the Soret coefficient~\cite{SegreEtAlPRE,Mixtures3,VailatiGiglio1,miJCP,miDynamics14}. Then
\begin{equation}\label{GrindEQ__4_7_}
\overline{p}_{{\rm NE}}^{w} =\frac{k_{{\rm B}} \overline{T}_{0} }{2\rho \nu D} A_{w} \overline{F}\; \overline{w}_{0}^{2} \left(1-\overline{w}_{0} \right)^{2} S_{T}^{2} L\left(\nabla T_{0} \right)^{2},
\end{equation}
where we have approximated the local temperature and concentration by their average values, $\overline{T}_{0}$ and  $\overline{w}_{0}$, in the center of the liquid layer. Experimentally, it may be more practical to study the NE fluctuation-induced pressure as a function of the distance $L$ at a fixed temperature difference $\Delta T=L\nabla T_{0}$. Substituting Eq.~\eqref{GrindEQ__2_11_} for the amplitude $A_{w}$ into Eq.~\eqref{GrindEQ__4_7_}, we obtain as our final expression for the NE pressure induced by the concentration fluctuations:
\begin{equation}
\begin{split}
\overline{p}_{{\rm NE}}^{w} =-&\frac{k_{{\rm B}} \overline{T}_{0}^{2} \left(\gamma -1\right)}{2\alpha \nu D} \frac{M^{3} }{M_{1}^{2} M_{2}^{2} }\\ &\times \left[\left(\frac{\partial ^{2} H^{{\rm E}} }{\partial x_{1}^{2} } \right)_{\mkern-7mu{p,T}} \hspace*{-12pt}-\frac{\rho c_{p,w} }{\alpha } \left(\frac{\partial ^{2} V^{{\rm E}} }{\partial x_{1}^{2} } \right)_{\mkern-7mu{p,T}} \right]\\&\times \overline{F}\; \overline{w}_{0}^{2} \left(1-\overline{w}_{0} \right)^{2} S_{T}^{2} L\left(\frac{\nabla T_{0} }{\overline{T}_{0} } \right)^{2}.\label{GrindEQ__4_8_}
\end{split}
\end{equation}

\begin{table*}\caption{Estimated NE fluctuation-induced pressures ($\overline{T}_{0} =298$~K, $\Delta T=25$~K)}\label{T1}
\begin{tabular*}{\textwidth}{ll@{\extracolsep{\fill}}cccc}
\toprule
&&$L = 10^{-6}$~m&$L = 10^{-5}$~m&$L = 10^{-4}$~m&$L = 10^{-3}$~m\\
\colrule
$\overline{p}_{{\rm NE}}^{w}$,&water+methanol$^\text{a}$&$+2\times10^{-1}$~Pa&$+2\times10^{-2}$~Pa&$+2\times10^{-3}$~Pa&$+2\times10^{-4}$~Pa \\
$\overline{p}_{{\rm NE}}^{w}$,&tetralin+$n$-dodecane$^\text{a}$&$+3\times10^{-1}$~Pa&$+3\times10^{-2}$~Pa&$+3\times10^{-3}$~Pa&$+3\times10^{-4}$~Pa \\
$\overline{p}_{{\rm NE}}^{w}$,&toluene+$n$-hexane$^\text{a}$&$+4\times10^{-1}$~Pa&$+4\times10^{-2}$~Pa&$+4\times10^{-3}$~Pa&$+4\times10^{-4}$~Pa \\
$\overline{p}_{{\rm NE}}^{w}$,&aniline+methanol$^\text{a}$&$-6\times10^{-1}$~Pa&$-6\times10^{-2}$~Pa&$-6\times10^{-3}$~Pa&$-6\times10^{-4}$~Pa \\
$\overline{p}_{{\rm NE}}^{w}$,&1-methylnaphtalene &$+17$~Pa&$+2$~Pa&$+2\times10^{-1}$~Pa&$+2\times10^{-2}$~Pa\\
&+$n$-heptane$^\text{a}$& \\
$\overline{p}_{{\rm NE}}^{T}$,&water~\cite{miPRE2014}&$+5\times10^{-1}$~Pa&$+5\times10^{-2}$~Pa&$+5\times10^{-3}$~Pa&$+5\times10^{-4}$~Pa \\
$\overline{p}_{{\rm NE}}^{T}$,&$n$-heptane~\cite{miPRE2014}&$+2\times10^{-1}$~Pa&$+2\times10^{-2}$~Pa&$+2\times10^{-3}$~Pa&$+2\times10^{-4}$~Pa \\
\botrule
\end{tabular*}
\flushleft $^\text{a}${\scriptsize Equimolar mixture}\\
\end{table*}

In Table~\ref{T1} we present some estimated NE fluctuations-induced pressures in a liquid layer with an average temperature of 298~K subjected to a temperature difference of 25~K. In addition, we compare the NE pressures $\overline{p}_{{\rm NE}}^{w} $, given by Eq. \eqref{GrindEQ__4_8_} from NE concentration fluctuations for two liquid mixtures, with NE pressures $\overline{p}_{{\rm NE}}^{T} $ from NE temperature fluctuations previously found for water and \textit{n}-heptane~\cite{miPRE2014}. Generally, the NE fluctuation-induced pressures in simple liquid mixtures are comparable to those in one-component liquids, as can be seen by comparing the NE-pressure values for toluene+\textit{n}-hexane with those for \textit{n}-heptane. The approximation of large Lewis and Schmidt numbers uncouples NE concentration and temperature fluctuations. However, temperature fluctuations are still coupled to the fluctuating Stokes equation~\eqref{GrindEQ__3_2_}, maintaining a structure similar to the one-component fluid case~\cite{BOOK,miIMT6}. As a consequence, in addition to pressures induced by NE concentration fluctuations, there will be pressures induced by NE temperature fluctuations, that in the large-Lewis -number approximation are given by the same expressions as those for one-component fluids~\cite{miPRL2,miPRE2014}, but with properties referred to those of the mixture. The corresponding order of magnitude is the same as the $\overline{p}_{{\rm NE}}^{T}$ for the examples shown in Table~\ref{T1}. In the parallel-plate configuration considered here, typical experimental separations are of the order of microns~\cite{BressiEtAl02,AntoniniEtAl,ZouEtAl}. The NE Casimir effect may make it possible to measure Casimir forces at larger length scales~\cite{miPRE2015}.

The data in Table~\ref{T1} show, as discussed in some previous papers~\cite{miPRL2,miPRE2014,miPRE2015}, that NE fluctuation-induced pressures are significantly larger than Casimir pressures induced in equilibrium fluids by critical fluctuations. The physical reason is that the intensity of both NE concentration and NE temperature fluctuations varies in the absence of boundaries with the wave number as $q^{-4}$, while critical fluctuations only vary as $q^{-2}$~\cite{Fisher64}. Since in liquid mixtures the NE Casmir effect is proportional to the square of the Soret coefficient, the effect can be further enhanced by selecting a liquid mixture with a large Soret coefficient~\cite{HartmannEtAl}. This is the reason why $\overline{p}_{{\rm NE}}^{w} $ is much larger in 1-methylnaphtalene+\textit{n}-heptane than in toluene+\textit{n}-heptane. We also see from Eq.~\eqref{GrindEQ__4_8_} that $\overline{p}_{{\rm NE}}^{w} $ is strongly related to the concentration dependence of the excess molar enthalpy and excess molar volume. Hence, as shown in Table~\ref{T1}, $\overline{p}_{{\rm NE}}^{w}$ can be either positive or negative.

\begin{figure}[t]
\begin{center}
\includegraphics[width=0.9\columnwidth]{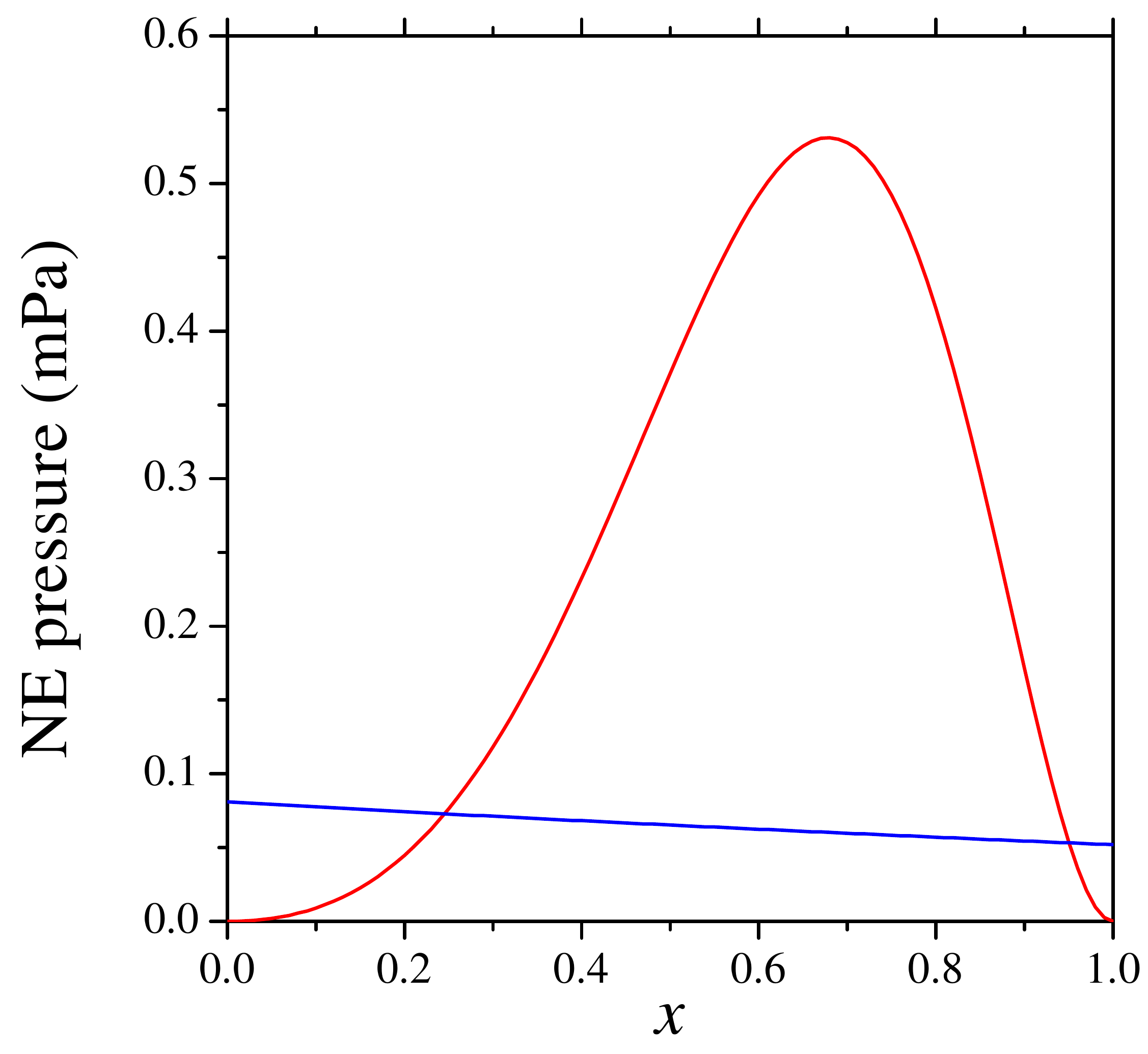}
\end{center}
\caption{(color on line). NE fluctuation-induced (Casimir) pressures as a function of the molar fraction $x$ of tetralin in a binary mixture with n-dodecane at atmospheric pressure, for $\Delta{T}=25~\text{K}$ and $L=1~\text{mm}$. Red curve represents the pressure induced by NE concentration fluctuations, Eq.~\eqref{GrindEQ__4_8_}. The blue curve represents the pressure induced by NE temperature fluctuations.}
\label{F1b}
\end{figure}

For the system tetralin+n-dodecane, all required thermophysical properties are available as a function of the concentration: excess enthalpy~\cite{LechterScoones} and volume~\cite{ParedesEtAl}, diffusion and Soret coefficients~\cite{GebhardtEtAl}, \emph{etc.} Therefore, for this particular system it is possible to study the dependence of $\overline{p}_{{\rm NE}}^{w}$ on the mixture concentration. In Fig.~\ref{F1b} we represent as a red curve the pressure induced by NE concentration fluctuations, calculated from Eq.~\eqref{GrindEQ__4_8_}, as a function of the tetralin mole fraction in the mixture, $x$. Temperature difference is $\Delta{T}=25~\text{K}$ and $L=1~\text{mm}$, corresponding to the right-most column in Table~\ref{T1}. We are also plotting in Fig.~\ref{F1b} the pressure induced by NE temperature fluctuations~\cite{miPRL2,miPRE2014} calculated for the mixture properties, so that it slightly depends on concentration. The total pressure induced by NE fluctuations will be the sum of the two curves in Fig.~\ref{F1b}.

We note that, due to their spatial long-range nature, the NE concentration fluctuations depend on gravity~\cite{VailatiGiglio1,miIMT6,miPRE2}. Essentially, gravity (buoyancy) suppresses the intensity of NE concentration fluctuations when it is parallel to the correspondingly induced stationary density gradient, while it further enhances NE fluctuations when it is antiparallel to this density gradient. The effects of buoyancy are more prominent the larger the spatial size of the fluctuation. Hence, gravity affects mostly fluctuations of small $q$, and competes with boundary effects in this wave number range. As a consequence, the NE pressures discussed in this paper will also depend on gravity. The effect of gravity on $\overline{p}_{{\rm NE}}^{T}$ has been elucidated in a previous publication~\cite{miPRE2014}. It turns out that the effect of gravity is modest when the liquid is far away from any hydrodynamic instability. Therefore, we expect that the effect will also be modest in liquid mixtures far away from any hydrodynamic instability. However, NE pressures will diverge at the onset of any convective instability~\cite{miPRE2014}.

Our result for the NE Casmir effect in liquid mixtures is fundamentally different from the Casimir-Soret effect discussed by Najafi and Golestanian~\cite{NajafiGolestanian}. Najafi and Golestanian have used a Langevin equation for a Goldstone mode to obtain an estimate for a Soret-like effect in a model system. They do not consider any mode coupling in the fluctuating mass-diffusion equation~\eqref{GrindEQ__3_1_}, but do account for the variation of the temperature in the expression \eqref{GrindEQ__3_3_} for the fluctuation-dissipation theorem when the mixture is subjected to a temperature gradient. They conclude that the inhomogeneity of the noise causes a thermophoretic force that is linear in the temperature gradient $\nabla T_{0}$. Fluctuations caused by the inhomogeneity of the noise terms in the fluctuating-hydrodynamics equations are of much shorter range than fluctuations induced by mode coupling~\cite{miPRE2015,miStatis}. As a consequence, Najafi and Golestanian find a force that is very strongly dependent on a molecular cutoff. In general one should expect several types of fluctuation-induced forces near walls, both from molecular origin and from long-ranged fluctuations. As pointed out in the literature~\cite{GambassiEtAl1,Krech1}, one normally identifies Casimir effects with those resulting from truly long-ranged fluctuations that induce forces that are independent of any molecular cutoff.

\section{DISCUSSION\label{S5}}

\begin{figure}[b]
\begin{center}
\includegraphics[width=0.9\columnwidth]{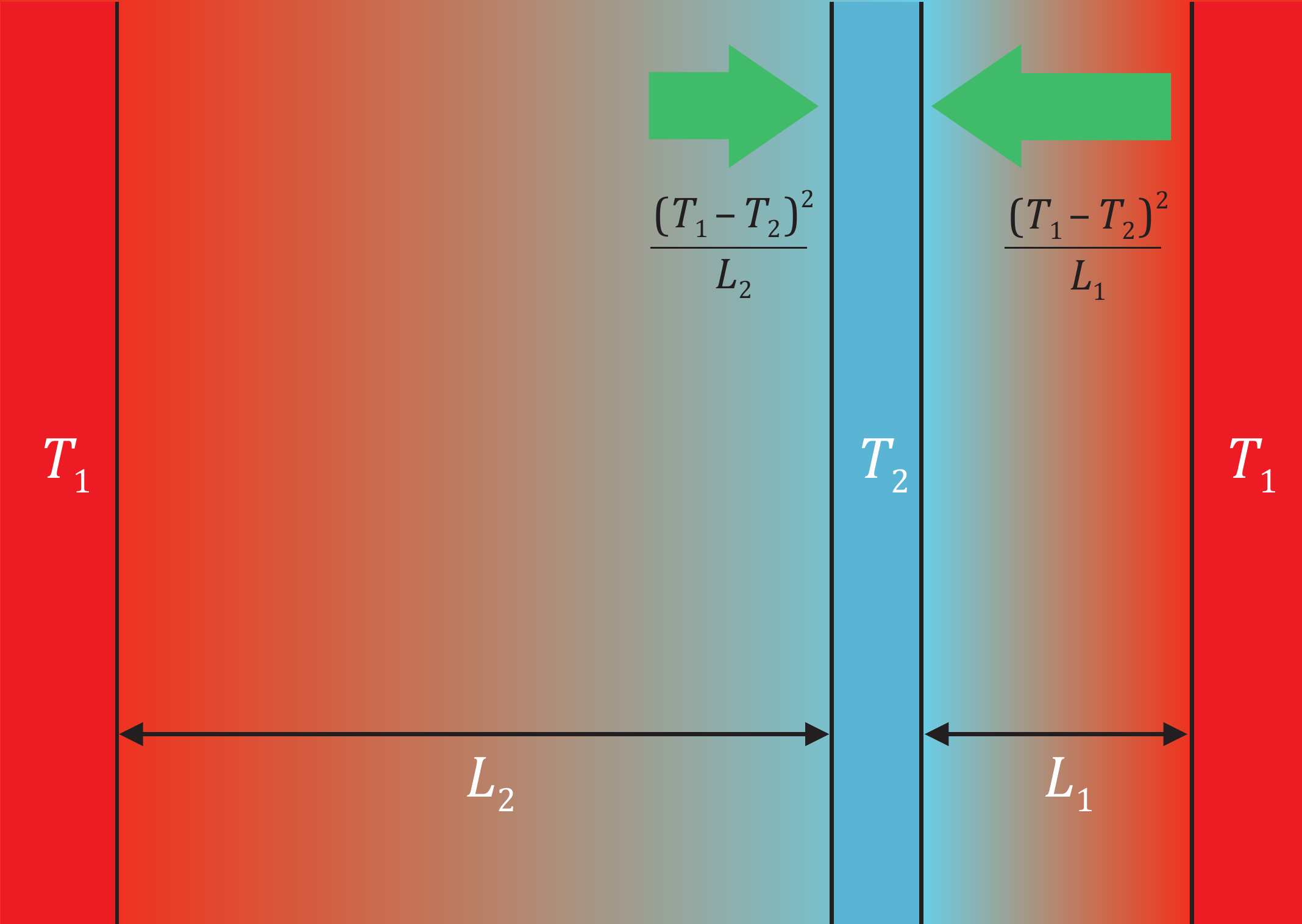}
\end{center}
\caption{Schematic illustration of a NE Casimir pressure $\overline{p}_{{\rm NE}} $ $>$ 0 on a plate with temperature $T_{2} $ located in a liquid between two walls with temperature $T_{1} $. For $\overline{p}_{{\rm NE}} $ $<$ 0, the plate would be pulled to the closest wall.}
\label{F2}
\end{figure}

We have shown that in liquid mixtures NE concentration fluctuations induce NE pressures that are proportional to the square of the concentration gradient $\nabla w_{0}$ and increase with the distance $L$. Like NE pressures induced by NE temperature fluctuations~\cite{miPRL2,miPRE2014,miPRE2015}, they are significantly larger than Casimir-like pressures induced in equilibrium fluids by critical concentration fluctuations. In liquid mixtures these NE fluctuation-induced pressures can be further enhanced by selecting a mixture with a large Soret coefficient, as illustrated in Table~\ref{T1}.

As a possible implication we may imagine a configuration where a thin plate with temperature $T_{2} $ is located in a liquid between two walls, both at a temperature $T_{1}$, as schematically shown in Fig.~\ref{F2}. When $\overline{p}_{{\rm NE}}>0$, the NE pressure will exert forces on the two sides of the inner plate proportional to $\left(\Delta T\right)^{2} /L_{1}$ and $\left(\Delta T\right)^{2} /L_{2}$. This will be the case either for a one-component liquid or a liquid mixture ($\overline{p}_{{\rm NE}}^{T} $ or $\overline{p}_{{\rm NE}}^{w} $). When $\overline{p}_{{\rm NE}}>0$ and $L_{1} \ne L_{2}$, the plate will experience a net force causing it to move away from the walls. Hence, the force needed to move this plate off center would be a measure of the NE Casimir force. In practice it may be difficult to maintain plates at a close distance parallel to each other~\cite{Parker92,BressiEtAl02,KlimchitskayaEtAl}. Hence, in studying Casimir forces, one commonly measures the force on a particle close to a surface~\cite{HertleinEtAl}. While a geometrical analysis of such a configuration becomes more complicated~\cite{Krech1,TrondleEtAl,LawEtAl2}, the physical principle remains the same.

NE Casimir forces are to be distinguished from thermophoretic forces ${\bf E}_{{\rm th}}$ on a particle in a liquid subjected to a temperature gradient~\cite{SchermerEtAl}:
\begin{equation} \label{GrindEQ__5_1_}
{\bf E}_{{\rm th}} =-6\pi RD_{{\rm th}} {\bf \nabla }T_{0} ,
\end{equation}
where $D_{{\rm th}}$ is its thermophoretic mobility in the liquid and $R$ the radius of the particle.  This thermophoretic force has been investigated experimentally by Regazzetti \textit{et al.} for silica particles with a radius $R=\SI{3}{\micro\meter}$ in a number of liquids, including water and \textit{n}-heptane~\cite{RegazettiEtAl}. The experiments were conducted in a liquid layer with $L=0.1$~m subjected to a temperature difference $\Delta T=25$~K. This is the reason why Table~\ref{T1} gives estimates for the NE pressures with $\Delta T=25$~K. From the experimental results of Regazzetti \textit{et al.}, we have earlier concluded that for silica particles in water $E_{{\rm th}} =280$~fN and in \textit{n}-heptane $E_{{\rm th}} =30$~fN~\cite{miPRE2014}. Most recently, Helden \emph{et al.}, by adopting a sophisticated optical technique that takes advantage of evanescent light after a total reflection~\cite{HeldenEtAl}, have directly measured the thermophoretic force experienced by polystyrene particles of \SI{2.5}{\micro\meter} radius in water. They report somehow smaller forces, of the order of $E_{{\rm th}} =50$~fN for a temperature gradient of \SI{0.14}{\kelvin \ensuremath{\cdot}\micro\meter^{-1}} 
(corresponding to $L\simeq 2\times10^{-4}$~m for $\Delta{T}=25$~K). On the other hand, we find from the information in Table~\ref{T1} that the fluctuations induced NE force $E_{{\rm NE}} \simeq \pi R^{2} \overline{p}_{{\rm NE}}^{T} =340$~fN for micrometer-sized particles in water and $E_{{\rm NE}} \simeq 120$~fN for micrometer-sized particles in \textit{n}-heptane~\cite{miPRE2014}. Clearly, our predicted NE Casimir forces are comparable to thermophoretic forces on particles.

However, the major difference is that thermophoretic forces are linear functions of the temperature gradient, while the NE Casimir forces are proportional to the square of the temperature gradient. This suggests two possibilities for detecting NE Casimir forces experimentally. One procedure would be to study the force on a particle as a function of the temperature difference $\Delta T$ to see whether the force has a component that depends on the square of the temperature gradient $\Delta T$~\cite{HeldenEtAl}. An even more direct indication of the presence of NE Casmir forces would be to change the direction of the temperature gradient in the experiments. While a thermophoretic force would change sign upon changing the direction of $\nabla T_{0}$, the NE force should reveal itself as a component that is independent of the direction of the temperature gradient $\nabla T_{0}$.

\section*{ACKNOWLEDGMENTS}

The authors acknowledge valuable discussions with Jeremy N. Munday of the University of Maryland. The research at the University of Maryland was supported by the U.S. National Science Foundation under Grant No. DMR-1401449. The research at UCM was funded by the Spanish State Secretary of Research under Grant No. FIS2014-58950-C2-2-P.

\appendix

\section{Statistical-Mechanical Derivation\label{A2}}
\renewcommand{\theequation}{A\arabic{equation}}

In statistical mechanics the pressure is given by the diagonal element of the microscopic stress tensor averaged over the \textit{N}-particle distribution function, $\rho _{N} $. Here we consider a two-component fluid in a non-equilibrium steady state (NESS) that is close to local equilibrium. We can then decompose $\rho _{N}$ into a local equilibrium part, $\rho_{{\rm LE}}$, and a part linear in the macroscopic gradients, $\rho_{\nabla}$. The explicit expression for the local equilibrium part is
\begin{equation} \label{A2_1_}
\rho _{{\rm LE}} =\frac{\exp [y{\rm \star }a]}{{\rm Tr\; exp}[y{\rm \star }a]}
\end{equation}
with $\{a\}=\{n_1, n_2 ,\mathbf g,e\}$ the set of microscopic conserved quantities, $\{y\}=\{\beta (\mu_1 - m_{1} u^{2} /2), \beta (\mu_2- m_{2}u^{2}/2), \beta {\bf u}, -\beta\}$ the macroscopic conjugate variables, while $y{\rm \star }a=\int d{\bf r}~ y({\bf r})~a\left({\bf r}\right)$ denotes an integration over space. In these expressions, $n_i$ is the number density of species $i=(1,2)$, $\mathbf g$ is the momentum density, $e$ is the energy density, $\beta =1/k_{{\rm B}} T$ is the inverse temperature, $\mu_i$ is the chemical potential of species $i$, $m_i$ is the mass of species $i$, and $\bf u$ is the fluid velocity with magnitude $u$.

In a non-equilibrium steady state of a two-component fluid with a chemical potential gradient of species $1$, but no velocity gradients, or temperature gradients, Liouville's equation gives for the gradient part of the \textit{N}-particle distribution function a time-dependent integral of the form~\cite{Zubarev}
\begin{equation} \label{A2_2_}
\rho _{\nabla } =-\int _{0}^{\infty }dt{\kern 1pt} \exp \left(-\mathcal{L}t\right)\rho _{{\rm LE}} ~ \widehat{{\bf J}}_{1} {\rm \star }\frac{\partial y_{1} }{\partial {\rm x}} .
\end{equation}
Here $\mathcal{L}$ is Liouville's operator, $\widehat{{\bf J}}_{1}$ is the part of the mass current of species $1$ that is orthogonal to the conserved quantities~\cite{Wood}, and $y_{1} =\beta \mu_1$. The pressure is defined as $1/d$ ($d$ being the spatial dimension) times the average of the trace of the microscopic stress tensor. For our purpose we can relate it to one of the diagonal elements that we denote by $J_{l}$~\cite{ErnstHaugeVanLeeuwen2}. The non-equilibrium or gradient part of the pressure can then be written as,
\begin{equation} \label{A2_3_}
\begin{split}
p_{{\rm NE}} \left({\bf r}\right)&=\left\langle J_{l} \left({\bf r}\right)\right\rangle _{{\rm NE}}\\& =-\int _{0}^{\infty }dt{\kern 1pt} \left\langle J_{l} \left({\bf r}{\it ,t}\right)\widehat{{\it J}}_{1} \left(0\right)\right\rangle _{{\rm LE}}  {\rm \star }\frac{\partial y_{1} }{\partial {\rm x}}.
\end{split}
\end{equation}
Here $\left\langle \right\rangle _{{\rm NE}}$ denotes a non-equilibrium ensemble average and $\left\langle \right\rangle _{{\rm LE}}$ denotes a local-equilibrium ensemble average. Generally, $p_{{\rm NE}} \left({\bf r}\right)$ is a local NE pressure depending on the position ${\bf r=}\left\{x,y,z\right\}$. Equation~\eqref{A2_3_} has the structure of a Green-Kubo expression for a transport coefficient, namely, an unequal time current-current correlation function, integrated over all times \textit{t}~\cite{Zwanzig65}. Note, however, that the currents in the integrand of this equation are different, unlike the current-current correlation functions for the usual Navier-Stokes transport coefficients~\cite{DorfmanKirkpatrickSengers}. Hence, the NE pressure originates from a cross Onsager-like effect, \textit{i.e., }a normal stress or pressure is caused by a chemical potential gradient.

Techniques to evaluate the long-wavelength, or hydrodynamic-mode, contributions to local-equilibrium correlation functions like Eq.~\eqref{A2_3_} have been developed by Kirkpatrick \textit{et al.}~\cite{KirkpatrickEtAl4,KirkpatrickEtAl3,KirkpatrickEtAl}, who extended  the methods of Ernst \textit{et al.}~\cite{ErnstHaugeVanLeeuwen0,ErnstHaugeVanLeeuwen,ErnstHaugeVanLeeuwen2}  to non-equilibrium steady states. In the large Lewis number limit, the leading contribution is
\begin{equation} \label{A2_4_}
p_{{\rm NE}} \left({\bf r}\right)=\frac{1}{2} \left(\frac {\rho}{k_{B}T} \right)^{2} {\left(\frac {\partial {\mu}}{\partial w}\right)}^{2}_{p,T}\left\langle J_{l,0} w_{0} w_{0} \right\rangle {\langle [\delta w(\mathbf r)]^2\rangle}_{NE}.
\end{equation}
Here $\mu =\mu_1-\mu_2$  and the subscript 0 of the phase variables in the statistical-mechanical average in Eq.~\eqref{A2_4_} indicate that they are at zero wave number. Finally, Eq.~(109) in Ref.~\cite{Wood} gives,
\begin{equation} \label{A2_5_}
A_w=\left(\frac {\rho}{k_{B}T} \right)^{2} {\left(\frac {\partial {\mu}}{\partial w}\right)}^{2}_{p,T}\left\langle J_{l,0} w_{0} w_{0} \right\rangle,
\end{equation}
with $A_w$ given by Eq.~\eqref{GrindEQ__2_8_}. We conclude that Eq.~\eqref{A2_4_} for $p_{{\rm NE}}$ is identical to the one given in Section~\ref{S2}.


\end{document}